# Metasurface-based Mueller Matrix Microscope

*Jiawei Zuo[1,2], Ashutosh Bangalore Aravinda Babu[1], Mo Tian[1,2], Jing Bai[1,2], Shinhyuk Choi[1,2], Hossain Mansur Resalat Faruque[1,2], Smitha S. Swain[3], Michael N. Kozicki[1], Chao Wang[1,4], Yu Yao[1,2]\**

[1]School of Electrical, Computer and Energy Engineering, Arizona State University, Tempe, AZ, USA, 85281

[2]Center for Photonic Innovation, Arizona State University, Tempe, AZ, USA, 85281

[3]School for Engineering of Matter, Transport and Energy, Arizona State University, Tempe, AZ, USA, 85281

[4]Center for Molecular Design and Biomimetics at the Biodesign Institute, Arizona State University, USA, AZ 85287

**\*Corresponding author:** yuyao@asu.edu

## Abstract

In conventional optical microscopes, image contrast of objects mainly results from the differences in light intensity and/or color. Muller matrix optical microscopes (MMMs), on the other hand, can provide significantly enhanced image contrast and rich information about objects by analyzing their interactions with polarized light. However, state-of-art MMMs are fundamentally limited by bulky and slow polarization state generators and analyzers. Here, we demonstrated the feasibility of applying metasurfaces to enable a fast and compact MMM, i.e., Meta-MMM. We developed a dual-color MMM, in both reflection and transmission modes, based on a chip-integrated high-speed metasurface polarization state analyzer (Meta-PSA) and realized high measurement accuracy for Muller matrix (MM) imaging. We applied our Meta-MMM to nanostructure characterization, surface morphology analysis and discovered birefringent structures in honeybee wings. Our meta-MMMs hold the promise to revolutionize various applications from biological imaging, medical diagnosis, material characterization to industry inspection and space exploration.

# Main

## Introduction

Besides intensity and spectrum, polarization is also a fundamental light property containing important information about objects that emit, reflect, transmit, or scatter light. Polarization microscopy can reveal the unique microscopic features of specimens due to scattering, emission, birefringence, etc. It has been widely used to provide microscopic information on film thickness, molecular structures, surface morphology, homogeneity, etc. [1-3]. Thus it is useful for chemical analysis [4, 5], biomedical imaging [6], cancer diagnosis [7, 8], space and industrial applications [9-11]. The Muller Matrix Microscope (MMM) is a natural evolution of the polarization microscope which enables quantitative characterizations of optical properties, such as linear birefringence, linear dichroism, depolarization, circular birefringence and circular dichroism of specimens [12]. In practice, the MMMs can be widely used in many applications such as biological and clinical research [13-18], fluorescence imaging[19, 20], agriculture and food industry[21, 22], industry imaging[23, 24], material science[25, 26].

However, conventional MMMs are bulky and slow because they generally rely on rotating linear polarizers (LPs) and quarter-wave plates (QWPs) in the polarization state generators (PSGs) and PSAs [27, 28]. These systems require taking a number of images (N≥16) using incident light of multiple different polarization states to obtain the complete MM information and are usually slow (1-2 min per MM with measurement error <2% [28]). Moreover, the system's measurement accuracy is sensitive to the mechanical positioning of the QWPs and LPs. Modulation-based techniques were developed to improve MM measurement accuracy and speed, such as utilizing liquid-crystal variable retarders (LCR)[29, 30], and Photo elastic modulators (PEM) [31, 32] to replace mechanically moving parts (>5 s per MM for LCR[30] and >320 ms per MM for PEM). However, complicated demodulation processes were required [29-32] for accurate MM measurement, and thus computation costly. Recently, spatial-division polarimeter arrays have been demonstrated for MM imaging[33] with fast measurement speed (~2.6s per MM image[33]), as their PSA can take complete measurement of polarization states at a single snapshot within 0.5 ms[33] and a smaller number of images (N≥4) are required for full MM measurement because. However, such a system requires two linear polarimetric imaging sensors and additional optical components, making the setup complex and expensive. So far, it remains challenging to realize MMMs with high speed, high accuracy, compact and simple configurations.

Here, we present a compact, dual-mode, dual-color MMM adopting a chip-integrated PSA based on metasurface devices, i.e., Meta-PSA. The Meta-PSA can simultaneously perform full-Stokes polarimetric detection for thousands of spatial points (readily scalable up to millions of points) in a single snapshot. We demonstrated the Meta-MMM in both transmission and reflection modes with high measurement accuracy for full-Stokes polarimetric imaging and MM imaging (MM measurement errors are about 1% to 2%) for red and green colors. Compared with the state-of-art MMM systems, our proposed Meta-MMM is featured with ultra-compact, high speed (~2s per MM image, limited by the CMOS imaging sensor), compact and simple system configuration. Furthermore, we applied the proposed Meta-MMM system to characterize nanostructured thin films, silver dendritic particles, and also discovered for the first time the optical birefringence in honeybee wings, suggesting its broad applications in material and structure characterization, industrial inspection, biological study, etc.

## Design concept

Figure 1a shows a schematic of the proposed Meta-MMM with a Meta-PSA. Unlike conventional PSAs using bulky optical components, the Meta-PSA is based on an ultra-thin metasurface-based microscale polarization filter array (MPFA) with a total thickness of the functional layers <640 nm. There are two ways to use the MPFA for polarization state measurement. One way is to use optical lenses to project the metasurface onto the focal plane of the imaging sensor [34, 35]. This method does not require a chip integration process, but the extra image projection lenses make the system bulky, complicated, and challenging for alignment. Another method is directly integrating the MPFA onto a imaging sensor[36]. This chip-integration method can lead to the most compact and mechanically stable system configuration. Here, we chose the chip-integration method and thus obtained an ultra-compact Meta-PSA directly integrated onto a CMOS imaging sensor (Fig. 1b).

The polarization state measurement using the chip-integrated Meta-PSA is carried out based on the spatial division polarization measurement concept. This method can enable very high-speed polarization and MM imaging because the polarization states of all imaging pixels are obtained at a single snapshot, and the measurement speed is ultimately limited by the imaging sensor. Figure 1c illustrates the device's configuration and working principle. The Meta-PSA consists of over 75,000 microscale metasurface polarization filters. In each super-pixel, there are four linear polarization (LP) filters ($P_1$ to $P_4$) and two metasurface circular polarization (CP) filters ($P_5$ to $P_6$) for full Stokes polarization states analysis with high accuracy. The LP filters, $P_1$-$P_4$, are based on vertically coupled double-layered Al gratings (VCDG) designed to transmit LP components polarized along 0°, 90°, 45°, and 135° respectively. For LP filter $P_1$ (Fig. 1d), the incident light polarized along *x*-axis can transmit through VCDGs with high efficiency, while light polarized along the y-axis is almost completely blocked, resulting in a very high linear polarization extinction ratio (LPER). The CP filters, $P_5$ and $P_6$, are based on multi-layered chiral metasurfaces transmitting right-handed circularly polarized light (RCP) and left-handed circularly polarized light (LCP), respectively. They are composed of a top Si metasurface, a silicon oxide ($SiO_x$) spacer layer, and a bottom layer of VCDG (Fig.1e). The Si metasurface layer consists of Si subwavelength nanogratings which are oriented at an angle of ±45° with respect to bottom VCDG. The Si metasurface is designed with structurally induced strong optical birefringence, i.e., phase difference of $\frac{\pi}{2}$ at red (at wavelength of ~630nm) and $\frac{3\pi}{2}$ at green (at wavelength of ~500nm), allowing the chiral metasurface functioning at dual operation wavelengths [37, 38]. For the LCP filter ($P_6$), incoming LCP light transmitting through Si metasurface is converted to LP polarized along *x*-axis and hence transmits through the bottom VCDG, while RCP light is converted to LP polarized along *y*-axis and hence mostly blocked by VCDG (Fig. 1e). Thus, LCP filters selectively transmit LCP light and block RCP light with high CP extinction ratio (CPER). The MPFA was fabricated by firstly patterning a thin layer of amorphous silicon (*a*-Si) with electron beam lithography (EBL) and inductively coupled plasma - reactive ion etching (ICP-RIE) of *a*-Si, followed by silicon oxide spacing layer deposition by sputtering. Next, a second layer alignment and patterning of VCDGs were achieved using EBL, followed by reactive ion etching (RIE) of $SiO_x$ and metal deposition of 80nm Aluminum (Al). Finally, the MPFA was integrated into the CMOS imaging sensor with the VCDG layer facing down on the imaging sensor, using a UV-bonding technique to form Meta-PSA (See detailed information in Methods). Note that in each super-pixel, we added another pair of CP filters (*P5'* and *P6'*) identical to $P_5$ and $P_6$ as back-up sub-pixels for CP filters in case of defects generated during nanofabrication. The SEM images of fabricated Meta-PSA are shown in Supplementary information (Fig. S1). We then evaluated the optical performance of Meta-PSA using setups as illustrated in Fig. S2 (efficiency measurement) and Fig. S4 (LPER and CPER measurement). The measurement results show that over 90% of the LP filters have LPER over 90 and over 85% of CP filters have CPER more than 15 (Fig. S5), while 80% of LP filters (*P1* to *P4*) have efficiency maintained over 40% and CP filters efficiency of ~30% (*P5* and *P6*) (Fig. S3). Assume the transmitted intensity of a super pixel forms an intensity vector $I_{cam} = (I_{P_1}, I_{P_2}, I_{P_3}, I_{P_4}, I_{P_5}, I_{P_6})^T$, any input polarization state $S_{in} = (S_0^{in}, S_1^{in}, S_2^{in}, S_3^{in})^T$ can be obtained by solving the equation below:

$$I_{cam} = A_{PSA} \cdot S_{in} \qquad (1)$$

Here, $A_{PSA}$ is the instrument matrix of one super-pixel of Meta-PSA. Detailed explanations of how to obtain $A_{PSA}$ theoretically and experimentally are presented in Supplementary information section 3. With the instrument matrix method, Meta-PSA can achieve high accuracy polarization detection with a measurement error of $S_1$, $S_2$, and $S_3$ less than 4% while maintaining high measurement speed (<34ms, fundamentally limited by the frame rate of the CMOS imaging sensor).

**Fig. 1: Meta-MMM and chip-integrated Meta-PSA design concept**

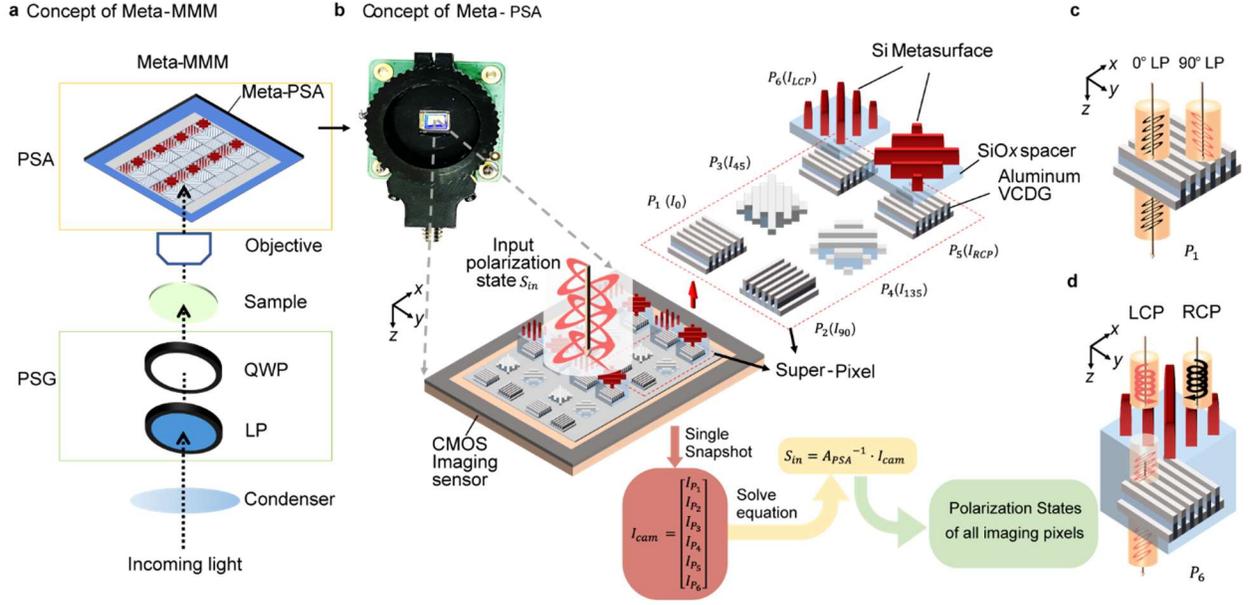

**a**, A 3D schematic of Meta-MMM. Polarized light generated by the PSG was transmitted through the sample and captured by the Meta-PSA. **b**, A Schematic of the Meta-PSA and conceptual demonstration of single snapshot full Stokes polarimetric imaging using the Meta-PSA. The Meta-PSA is composed of MPFA integrated into a commercial CMOS imaging sensor (Sony IMX477). $P_1$-$P_4$ denotes the microscale LP filters with transmission axes at 0° 90° 45° 135°, respectively, and $I_0$, $I_{45}$, $I_{90}$, $I_{135}$ denote corresponding transmitted intensity. $P_5$, $P_6$ denote microscale chiral metasurfaces transmitting RCP and LCP, respectively and $I_{LCP}$, $I_{RCP}$ denote the corresponding transmitted intensity. Top left inset: A photo of the Meta-PSA, the whole device (including the PCB board) has an area of less than 4×4 cm$^2$. **c**, A 3D schematic of VCDG LP filter ($P_1$) transmitting LP light with electric field vector oriented along $x$-axis (0° LP) while completely blocking LP light with electric field vector along $y$-axis (90° LP). **d**, 3D illustration of LCP filter ($P_6$) transmitting LCP light and blocking RCP light.

## Meta-MMM system configuration and performance

In this section, we first introduce the system configuration of the demonstrated Meta-MMM, then present the calibration process and the system performance evaluation results. Figure 2a shows a photo and a schematic of the Meta-MMM in transmission mode. The input polarization states were generated by a PSG composed of a rotating LP and QWP. The transmitted polarization image of the specimen was measured using Meta-PSA and a zoom lens system with tunable magnification (×1.16 to ×28, see Methods for more details of zoom lens system). A beam splitter was used for selecting the incoming light from transmission or reflection mode. Note that for simplicity, the setup for reflection mode measurement is not shown in Fig. 1a. A complete schematic of the Meta-MMM, including both transmission and reflection mode, is included in the Supplementary information (Fig. S6).

Figure 2b shows the flowchart for polarization measurement of the specimen, with polarization state $S_{in} = (S_0^{in}, S_1^{in}, S_2^{in}, S_3^{in})^{\mathrm{T}}$ generated by PSG as input. Input polarized light $S_{in}$ firstly transmitted through the specimen (MM written as $M_{s,T}$), the transmitted polarization state can thus be written as equation below:

$$S_{s,T} = M_{s,T} \cdot S_{in} \tag{2}$$

Next, polarized light transmits through beam splitters and lenses ,etc. before being captured by the Meta-PSA, this process can be described by the equations below:

$$S'_{s,T} = M_{sys,T} \cdot S_{s,T} \tag{3}$$

$$M_{sys,T} = M_{zoom\ lens} \cdot M_{bs\_T} \cdot M_{Objective} \tag{4}$$

Where $S'_{s,T}$ is the actual polarization states being captured by Meta-PSA, $M_{s,T}$ is the MM of the specimen in transmission mode; $M_{Objective}$ is the MM of the objective lens, $M_{bs\_T}$ is MM of beam splitter in transmission mode and $M_{zoom\ lens}$ is the MM of the Zoom lens, thus $M_{sys,T}$ is the MM for the complete Zoom lens system accounting for the combined polarization effect of the optical components mentioned above. Combining Eq.1 to Eq.4 above, the relationship between transmitted intensity vector $I_{cam}$ and input polarization state $S_{in}$ can be written as Eq.5 below:

$$I_{cam} = A_{sys,T} \cdot M_{s,T} \cdot S_{in} \quad (5)$$

$$A_{sys,T} = A_{PSA} \cdot M_{sys,T} \quad (6)$$

Where $A_{sys,T}$ is the instrument matrix of the whole system in transmission mode, accounting for the polarization effects of the Zoom lens system and the instrument matrix of Meta-PSA. Therefore, polarization state transmitted through the specimen ($S_{s,T}$) can be obtained using the Eq.7 below:

$$S_{s,T} = A_{sys,T}^{-1} \cdot I_{cam} \quad (7)$$

In addition, $M_{s,T}$ can also be obtained by measuring a number ($N \geq 4$) of polarization states using Meta-PSA (Fig. 2c), followed by solving the Eq.8 below (see Methods for details):

$$I_{cam,6 \times N} = A_{sys,T} \cdot M_{s,T} \cdot S_{in,4 \times N} \quad (8)$$

$$I_{cam,6 \times N} = [I_{cam}^1, I_{cam}^1 \ldots, I_{cam}^N] \quad (9)$$

$$S_{in,4 \times N} = [S_{in}^1, S_{in}^2, \ldots, S_{in}^N] \quad (10)$$

Where $S_{in,4 \times N}$ represents the Stokes parameters of N ($N \geq 4$) input polarization states generated by the PSG, $I_{cam,6 \times N}$ is the corresponding intensity matrix measured by the Meta-PSA. Theoretically, we need to measure at least four input polarization states (N=4) and make sure the rank of matrix $S_{in,4 \times N}$ to be 4 so that it is invertible to ensure that one can solve Eq.8 to obtain $M_{s,T}$, given the pre-calibrated system instrument matrix $A_{sys,T}$. In experiment, we chose $N$ to be 6 to ensure a stable and accurate measurement of $M_{s,T}$, to guarantee a high MM measurement accuracy. According to Eq. 8, one needs to perform calibration to obtain $A_{sys,T}$ prior to actual MM measurement. During the calibration process, the specimen was removed from the sample holder, and a number ($N=10$) of pre-known polarization states $S_{in,4\times N}$ generated by PSG was captured by the Meta-PSA in sequence to obtain intensity matrix $I_{cam,6\times N}$, $A_{sys,T}$ was then calculated by solving Eq.8 (see Methods for more details).

For reflection mode, we assume the recorded intensity of a super-pixel with a number of ($N\geq 4$) input Stokes parameters forms a matrix $I_{cam,6\times N}$, the relationship between MM of specimen in reflection mode ($M_{s,R}$) and $I_{cam,6\times N}$ can be written as:

$$I_{cam,6\times N} = A_{sys,R} \cdot M_{s,R} \cdot M_{BS,R} \cdot S_{in,4\times N} \qquad (11)$$

Where $A_{sys,R}$ is the instrument matrix of the system in reflection mode, $M_{BS,R}$ is combined MM of the objective and beam splitter (reflection mode). Therefore, $M_{s,R}$ can be obtained by measuring a number ($N\geq 4$) of polarization states using Meta-PSA followed by solving Eq.11 (see Methods for details). For reflection mode, $N$ is also chosen to be 6 to improve measurement accuracy and stability. Again, systematic calibrations for the Meta-MMM in reflection mode are needed to increase system measurement accuracy. One major difference between reflection and transmission modes is that $M_{BS,R}$ changes input polarization state ($S_{in}$) generated before it is incident onto the specimen, as shown in Eq.11. Therefore, $M_{BS,R}$ was measured separately before the instrument matrix calibration process. Detailed measurement methods and corresponding results of $M_{BS,R}$ (green and red color) are provided in Supplementary Information (Fig. S7). We then performed calibration process, which is similar to that for transmission mode to obtain the instrument matrix in reflection mode ($A_{sys,R}$). The flow chart and detailed discussion for the calibration process is included in the Supplementary information (Fig. S8) and Methods section.

Next, we performed polarization state detection to evaluate system performance in both transmission and reflection modes. For polarization measurement accuracy analysis, we measured 16 arbitrary polarization states in transmission mode using Meta-PSA at green (480 to 520nm) and red color (630 to 670nm), respectively, and compared them with results measured by traditional PSA (see Methods), as shown in Fig.2d. Stokes parameter measurement error analysis was then performed to evaluate the measurement results (Methods). The Mean Absolute Error (MAE) for $S_1$, $S_2$, and $S_3$ are less than 2.8% for transmission mode. Measurement results of input Stokes parameters for reflection mode with a silver mirror as specimen are presented in Supplementary Information (Fig. S9). The MAE for $S_1$, $S_2$, and $S_3$ are less than 3.7% for reflection mode. The MAE of $S_1$, $S_2$, $S_3$ in different colors can be found in Supplementary information (Table S3).

For MM imaging accuracy analysis, we measured the MM of a commercial linear polarizer (LPVISE100-A, Thorlabs) with different orientations in transmission mode and performed MM measurement error analysis (see Methods). The measured MM agrees well with theoretical MM values with MAE of 1.6% for red color and 2.1% for green, respectively, as shown in Fig. 2e. Then we measured the MM of a silicon wafer ($M_{Si\ wafer}$) in reflection mode, the MAE of the measured $M_{Si\ wafer}$ are 1.1% and 2.1% for red and green color, respectively. Detailed measurement results of $M_{Si\ wafer}$ is included in Supplementary Information section 7 (Table S4 and Fig. S10 for red and Table S5 and Fig. S11 for green).

So far, we have achieved a high accuracy full Stokes polarimetric imaging with an average error of 2~4% for transmission and reflection modes. We measured the MM of the linear polarizer, and silicon wafer, the measurement result shows MM measurement error is ~2% for both transmission and reflection modes. The measurement speed of the Meta-MMM is <34 ms per polarimetric image (limited by the speed of the CMOS imaging sensor), and the measurement speed of MM imaging is ~2s per MM, mainly limited by the rotation speed of QWP and LP in PSG. We attribute measurement error to the following major factors and plan to take corresponding measures to further improve the measurement accuracy of the system in future work. First, the chromatic dispersion of the chiral metasurface and the dispersion of the optical components in the optical systems has led to a wavelength dependent instrument matrix of the system. Thus, the final measured instrument matrix was an average over the wavelengths, leading to an accumulative measurement error over 40nm bandwidths. One way of overcoming such an issue is using narrow band light source such as narrow-band LED for better accuracy measurement. Meanwhile, we will also explore polarization filter designs, esp. CP filters, with smaller wavelength dispersion via dispersion engineering of metasurface. Secondly, the bonding process of MPFA onto CMOS imaging sensor unavoidably leaves a gap between the imaging sensor and the microscale polarization filters, thus resulting in crosstalk between adjacent subpixels. In the future, we plan to utilize foundry fabrication for customized CMOS imaging sensor with predefined alignment markers and realize direct integration of MPFA onto CMOS imaging sensor to reduce the vertical gap size and minimize crosstalk between adjacent pixels.

**Fig. 2: Full Stokes polarization detection of polarization states and Mueller matrix measurement.**

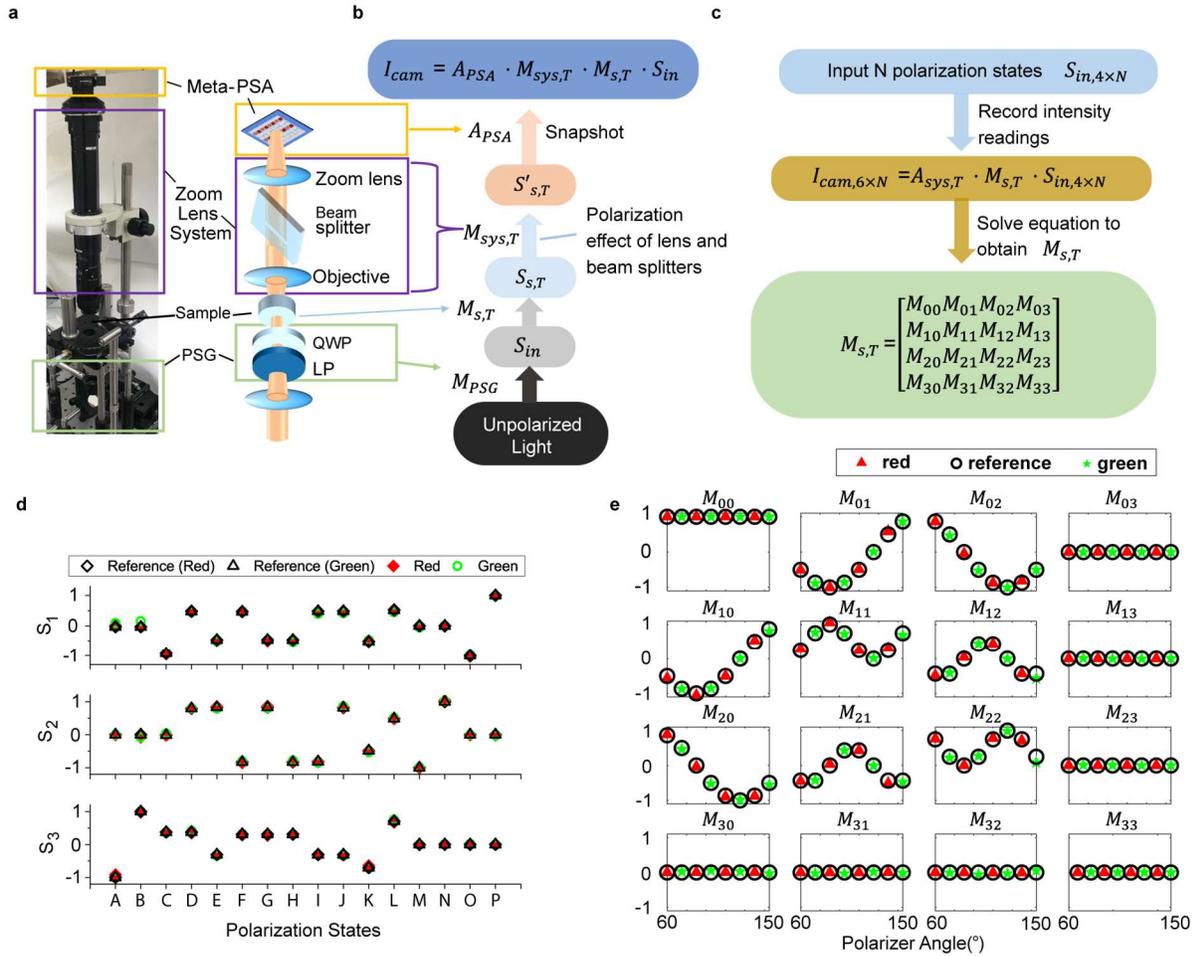

**a**, Photograph of meta-MMM showing Zoom lens system as objective, polarization state generator (PSG) in transmission mode, lens sets, and Meta-PSA. The corresponding field of view (FOV) and microscopic imaging resolution are 1.16mm×0.82mm, 3.66μm at the lowest magnification (×1.16) and 96 μm×67.9 μm, 0.83 μm at the highest magnification (×28), respectively. **b**, Flowchart for Stokes parameter measurement using Meta-MMM (transmission mode) based on the Instrument matrix method. **c**, Flowchart for MM measurement using Meta-MMM (transmission mode) based on the Instrument matrix method. **d**, Transmission mode full Stokes parameter measurement results of 16 arbitrary polarization states under 630 to 670nm (red square) and 480 to 520nm (green circle) input, respectively. Black hollowed square: theoretical values of Stokes parameters as a reference, red color; Black hollowed triangle: theoretical values of Stokes parameters as a reference, green color; red square: Measured results of Stokes parameter for red color; Green hollowed circle: Measured results of Stokes parameter for green color. **e**, Transmission mode MM measurement of a standard linear polarizer at different polarization axis angles. Black hollowed circle: theoretical values of a linear polarizer as reference. Red triangle: measured MM component results with red color input (630 to 670nm); green star: measured MM component results with green color input (480 to 520nm).

## Characterization of thin film structures

Muller matrix measurement has been widely used to characterize thin film structures, unveiling their birefringence, depolarization, scattering, and chiral properties. Here, we apply Meta-MMM to characterize an optical metasurface (OM) thin film structure with artificial birefringence designed for polarization control [37]. Studying the linear and circular retardance or diattenuation properties of optically thin artificial metamaterials[39] can facilitate the development of novel flat optical devices with high performance and low cost for numerous applications, such as metalens and plasmonic metasurfaces for polarization detection and polarimetric imaging [34, 40-42], etc. The microscopic and SEM images of Si metasurface composed of subwavelength Si nanogratings are shown in Fig. 3a. The Si metasurface exhibit strong optical linear birefringence and can function as a linear retarder [37] with the fast optical axis along U axis and slow axis along V axis (Fig. 3b). For incident light linearly polarized along *x*-axis, the transmitted light is converted into RCP at wavelengths around 500 nm (green light) and LCP at wavelength around 630 nm (red light). This is because the phase difference $\Delta\varphi$ between U (fast) and V (slow) axis are $\frac{3}{2}\pi$ at 500 nm and $\frac{1}{2}\pi$ at 630 nm (Fig. 3c). We used our meta-MMM to obtain the MM images of the Si metasurface and extracted the linear retardance from MM images (Fig. 3d) using Lu-chapman MM decomposition method[12] (see Methods). All 16 MM images of the Si metasurface for red and green incident light are shown in Supplementary Information (Fig. S12 for red, Fig. S13 for green). The spatially averaged linear retardance of Si metasurface on the left side (Figure 3d) are 0.436 $\pi$ in red (wavelength from 630 nm to 670 nm) and -0.557$\pi$ in green (wavelength from 480 nm to 520 nm), respectively, which agrees well with simulation results. In addition, we compared measurement results of polarization conversion obtained by Meta-MMM with results taken by a traditional PSA using rotating linear polarizer and quarter waveplate to evaluate measurement accuracy. Traditional methods and experimental setup for measuring polarization state transmitted through Si metasurface are discussed in detail in Supplementary information (Fig. S14). The degree of circular polarization (DOCP) images of light after passing the Si metasurface taken by Meta-MMM using 0° LP as input is shown in Fig. 3e. We obtained the spatially averaged value of DOCP image (spatial resolution:1.45µm) (left in Fig. 3e), i.e., -0.872 for red color (wavelength from 630 nm to 670 nm) and 0.813 for green (wavelength from 480 nm to 520 nm), respectively. We also measured the DOCP of light transmitted through the Si metasurface using the traditional PSA as shown in Fig.3f. The averaged converted DOCP values at red and green colors are -0.885 and 0.805, respectively. Thus the extracted measurement error for DOCP is less than 2%. Such high accuracy single-shot polarimetric imaging and high speed MM imaging of nanostructured thin film structures confirms the great potential for applications of Meta-MMM in material science and nanophotonic research.

**Fig.3: Characterization of optical birefringence in Si metasurface thin film structures**

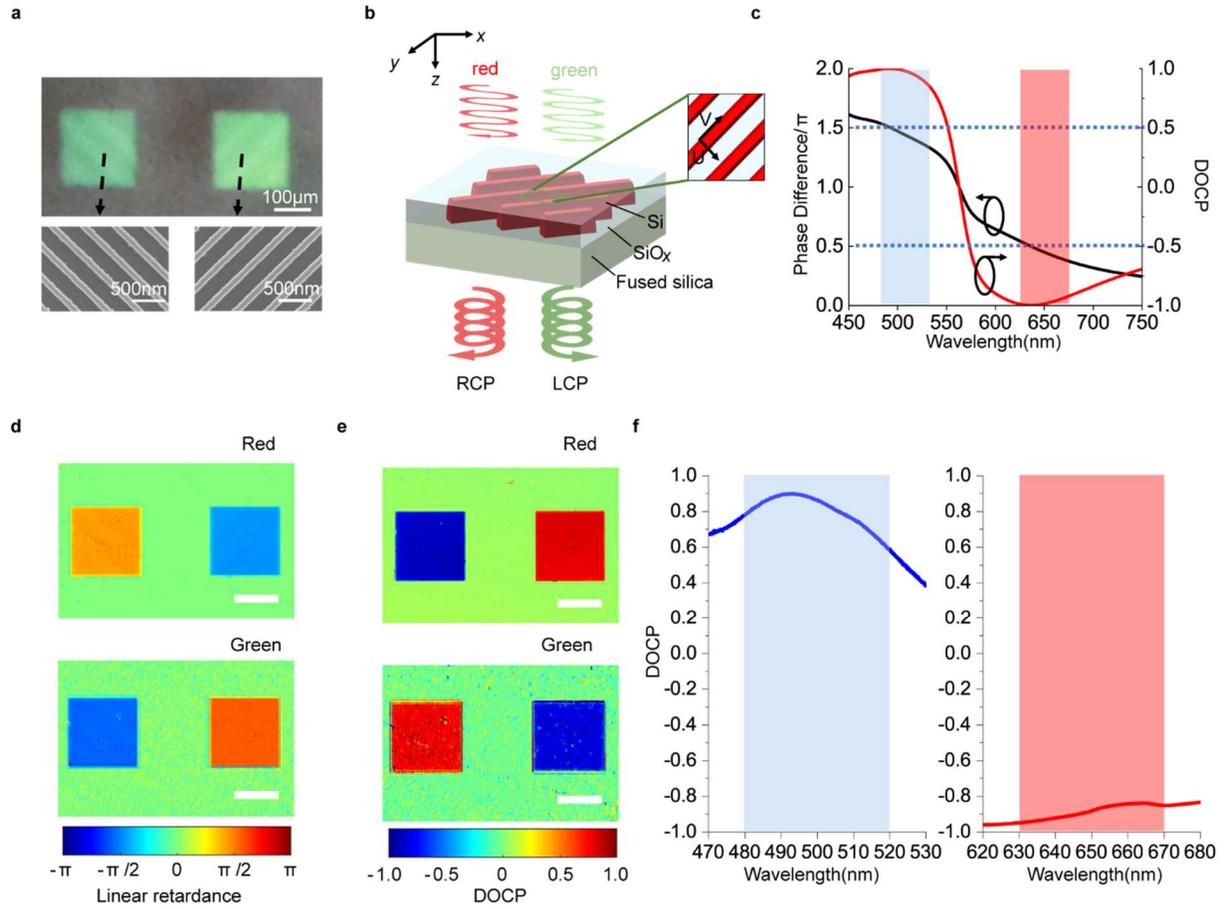

**a**, Optical microscope (top) and scanning electron microscope (SEM) (bottom) images of the fabricated Si metasurface structures. **b**, A 3D schematic of the Si metasurface. The width, period, and thickness of Si nanogratings are ~100nm, 297nm, and 130nm, respectively. **c**, Left axis: Simulated phase difference between fast (U) and slow(V) axes. Right axis: converted DOCP for LP input polarized along the *x*-axis. **d**, Linear retardance image of Si metasurface for red (wavelength from 630 nm to 670 nm) and green (wavelength from 480 nm to 520 nm) input, respectively. Scale bar: 100μm. Image Magnification: ×10. **e**, DOCP image measured by polarization microscopy under red (wavelength from 630 nm to 670 nm) and green (wavelength from 480 nm to 520 nm) input, respectively. Scale bar: 100μm. **f**, The DOCP of transmitted light measured by PSA using the traditional method at red (wavelength from 630 nm to 670 nm) and green color (wavelength from 480 nm to 520 nm) respectively.

**Characterization of complex metallic structures: silver dendrites**

Polarization and Muller matrix images are also useful for characterizing and analyzing materials with different surface morphology and scattering properties. In this work, we applied Meta-MMM to inspect the material properties of silver dendrites. Dendritic silver nanoparticles (AgNPs) are a type of dendrite-shaped conjugated metallic particles that can be fabricated via electrochemical[43], photochemical[44] methods, etc. The porous and metallic material property of these silver dendrites allow them to be optically scattering and absorptive, making them suitable for chemical sensing[45] and chemical catalysis[46]. In addition, the growth process of silver dendrites follows random Brownian motion; thus, the fractal shape of formed silver dendrites is always unique and this, couples with their nanostructure, makes them intrinsically unclonable and therefore ideal for physical tags for anti-counterfeiting applications[44, 47]. Despite a tremendous amount of effort in the chemistry and nanofabrication of silver dendrites, studies on the polarization effect introduced by unique material properties of silver dendrites are still at an early stage and of great interest, as polarization properties of silver dendrites induced by optical scattering can potentially reveal its surface morphology. Figure 4a shows a zoomed-in image (×2) of silver dendrite grown by the electrochemical method [48] (see Supplementary Information section 9 for details of the Silver based dendritic pattern fabrication process). The conjugated silver dendrites are typically a few μm thick and consist of silver nanoparticles that are ~ 30 to 50 nm in diameter, as shown in Fig. 4b. A lot of empty space exists between the silver nanoparticles, leading to a highly porous structure. As a comparison, we fabricated a thin film structure made of silver with the same patterns but a smooth surface using UV lithography, silver deposition, and lift-off processes, as illustrated in Fig. 4c. The image processing method for copying the fractal shape of silver dendrites is discussed in detail in the Supplementary Information (Fig. S15 and related discussions). Figure 4d shows a zoomed-in image (×2) of the fake sample. The fabricated silver structures have a continuous and smooth surface profile with grain size at the scale of ~30-50 nm, as shown in the SEM image in Fig.4e. We took polarization and MM images using the Meta-MMM in reflection mode with red light illumination (wavelength range 630nm to 670nm). Figure. 4f shows the intensity and DOCP images for the authentic silver dendrites (top) and Fig. 4g shows the fake ones (bottom) under RCP light illumination. The intensity images of the two samples are quite similar. Yet, their DOCP images are easily distinguished because the silver dendrites have a strong depolarization nature due to their porous surface morphology, while the silver thin films in the fake sample have optically smooth surfaces and thus reflect light without reducing the DOCP. The depolarization images of both samples (Fig. 4h) were obtained by taking MM images and performing MM decomposition using the Lu-chapman MM decomposition method[12]. The depolarization of silver dendrites is indeed very high due to materials scattering and absorption, while the depolarization of the fake sample is close to 0. The measured MM images of the silver dendrites and fake sample can be found in Supplementary Information (Fig. S16 for silver dendrites and Fig. S17 for fake sample). This study suggests the potential application of polarimetric imaging for dendrites characterization and detection to authenticate the true (electrochemical) patterns and avoid sophisticated copying/counterfeiting, as the fractal shape can possibly be duplicated via image algorithms (as we did for the fake sample), whilst the duplication of unique surface morphology of the silver dendrites is much harder. Similarly, such surface morphology analysis can also be applied for other applications such as industry inspection[49], solar thermal energy[50] and, catalyst [51], etc.

**Fig. 4: Characterization of surface morphology in metallic structures.**

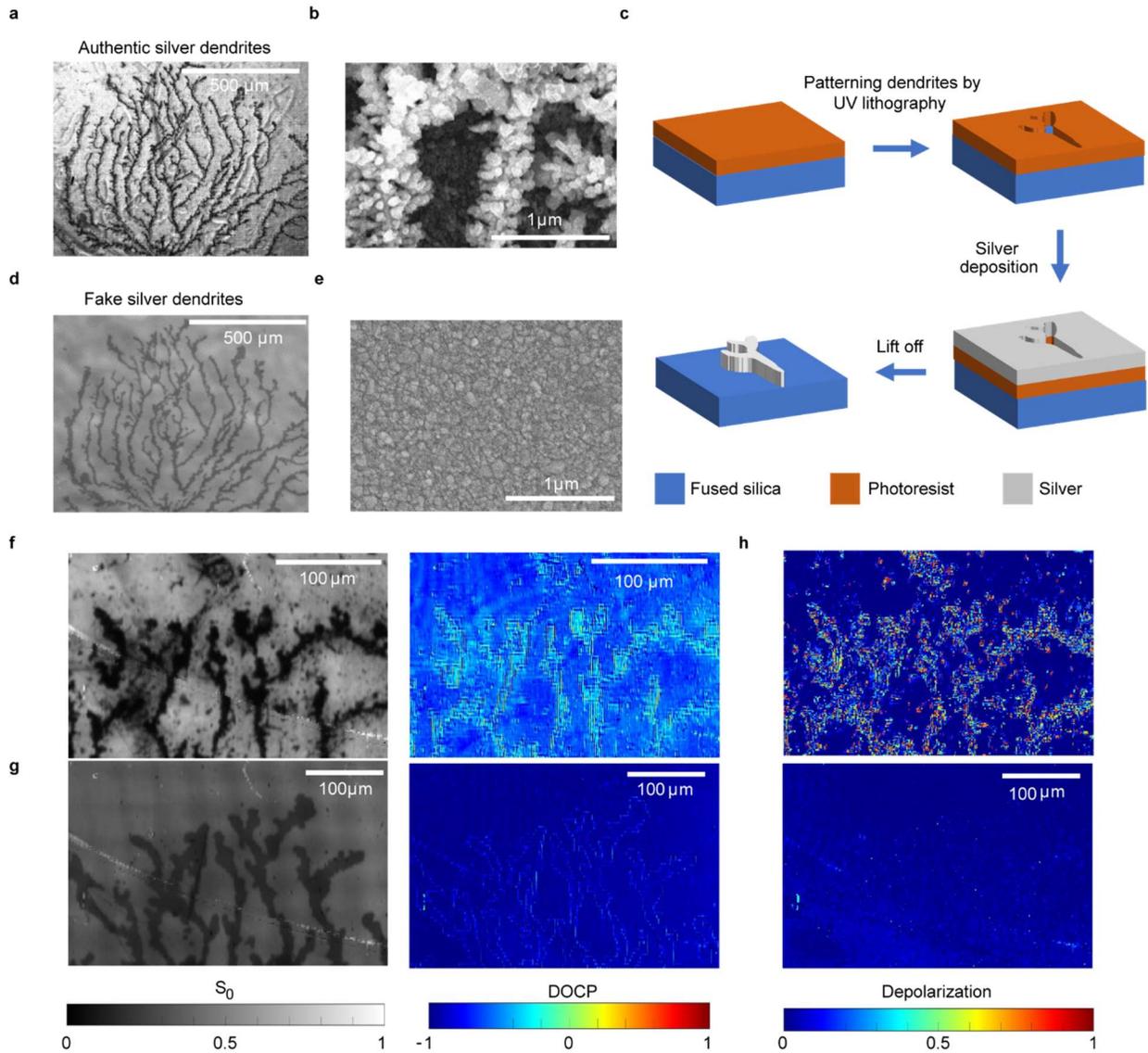

**a**, Zoomed-in image (×2) of grown silver dendrites. **b**, Scanning electron microscopic (SEM) image of grown silver dendrites. **c**, Fabrication flow chart of faking silver dendrites by UV lithography and metal deposition. A thin layer of Photoresist was spin-coated on the fused silica substrate, the faked silver dendrite was then patterned using a UV laser writer, followed by Ag thermal deposition and lift-off. **d**, Photo, and microscopic picture of fake silver dendrites. A sample photo is taken with white paper under the substrate to increase contrast. **e**, SEM image of duplicated silver dendrites by UV lithography. **f, g**, Full Stokes polarization image of grown silver dendrites and fake silver dendrites under red color input respectively. **h**, Depolarization image of grown and fake silver dendrites derived from their MM image under red color input.

**Imaging of biological samples: discovery of birefringence in honeybee wings**

Polarization information of biological samples has been extensively studied in biology and clinical research. For example, the linear birefringence was applied as an importance index for diagnosing cancer[6] and studying the composition of potato starch[52]. Moreover, linear or circular dichroism was used to study insects' behaviors and communications [53, 54], and plants' growth [55]. The polarization vision of honeybees has been studied extensively[56], and scientists have observed various shreds of evidence suggesting the use of polarization information by honeybees in their social activities [57]. Here, we investigated the wing of a honeybee using the Meta-MMM and revealed an optically birefringent structure on the honeybee wings for the first time in literature to our best knowledge. Figure 5a shows the zoomed-in images (×4 for the middle and ×28 for the right) of the hind wing of a honeybee, which is composed of wing cells and veins. We took MM images of one of the vein joints in transmission mode with red light illumination (wavelength from 630 nm to 670 nm). Figure 5b shows the DOCP images of vein joints under LP input light with polarization axis along 0°,45°,90°, and 135°, respectively. When the input light is polarized along or orthogonal to the vein joints, the transmitted light shows a small DOCP. When input light is ~±45° with respect to the vein joints, the transmitted light exhibits a DOCP value of ~±0.2. Such DOCP response suggests the tissues connecting bee wing cells and the vein joints have linear birefringence. The optical fast axis of vein joints is along the joint length direction, as shown by the arrows drawn in Fig. 5c. The linear retardance value of the vein joints extracted from the MM is shown in Fig. 5d, revealing an advanced phase difference of the vein joints with a value of

~$0.17\pi$ to $0.19\pi$ along the vein joints compared to the wing cells. We also obtained DOCP and extracted linear retardance images in green color (wavelength from 480 nm to 520 nm) and observed similar phenomena as described above, where linear retardance of ~$0.17\pi$ to $0.19\pi$ exists mainly at cells connecting the vein joints and wing cells, revealing a broadband optical birefringence characteristic of the vein joints. A complete set of polarization images, MM images, and linear retardance images in green color (wavelength from 480 nm to 520 nm) can be found in Supplementary information (Fig. S18, S19, S20). Our study suggests that besides polarization vision, Using MMM to study other body parts in bees might be useful for fully understanding how they make use of polarized light. Furthermore, our Meta-MMM could provide ultra-compact and high-speed solutions for obtaining polarization and MM images of other biological samples for tissue analysis[13], cancer diagnosis[58], and study of insects[56], plant growths[55], etc.

**Fig. 5: Full Stokes polarimetric images and Mueller matrix microscopic images of honeybee hind wings.**

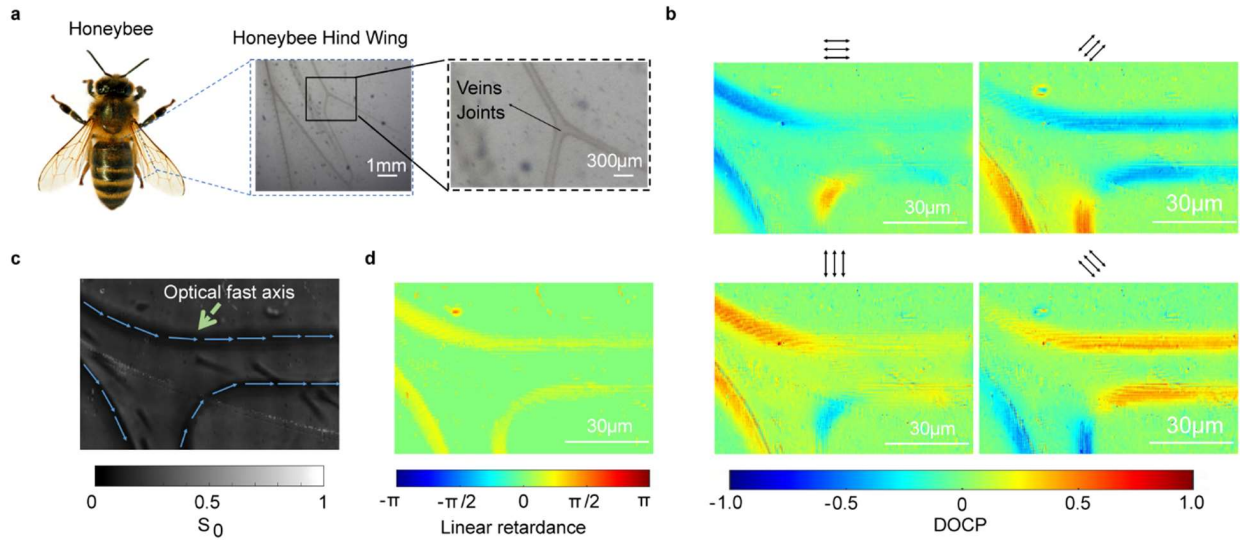

**a**, Microscopic image of honeybee hind wing vein joints. **b**, DOCP images of honeybee vein joints with LP input polarized along 0°,45°,90°,135°respectively, polarization images were taken under red light illumination (wavelength range from 630 to 670 nm). **c**, Illustration of the optical fast axis of the honeybee wing joints. **d**, Linear retardance extracted from MM of honeybee wing joints under red light illumination.

Conclusion

In conclusion, we have demonstrated polarimetric imaging and Muller matrix imaging microscopy based on metasurface devices with highly improved measurement speed, system compactness, and high measurement accuracy. By introducing a Meta-PSA to replace the conventional bulk and slow PSA, we significantly reduced the system complexity and achieved high-speed and full Stokes polarimetric imaging at a single snapshot. We proved that the metasurface MMM can enable highly accurate MM measurement (error 1%~2.1%) for both transmission and reflection modes with record speed (2s per MM image). We also explored the applications of our meta-MMM in characterization of nanostructured thin films (e.g., metasurfaces with artificially introduced birefringence), metallic structures with different surface morphology (e.g., silver dendrites and evaporated silver thin films), and discovered for the first time the birefringent structures in honeybee hind wings. Our research shows that metasurface devices can overcome limitations of conventional bulk optics and enable Meta-MMM has with ultra-compact and robust system configuration, broadband wavelength coverage, high measurement accuracy and speed. Considering the great design flexibility and ultra-compact form factor of metasurface devices, there are much room for device and system innovations based on metasurfaces to further improve polarization and Muller matrix microscope systems to be more compact (or even portable), faster, more accurate and cheaper, which are highly desirable for industrial inspection, biological study, clinical research, and space applications.

## Methods

*Simulations*. Lumerical Inc. Finite-Difference Time-Domain (FDTD) solver was used for full wave simulation to calculate the transmission efficiency and phase delay of the Si metasurface. The optical refractive index of *a*-Si thin film deposited by PECVD was measured by UV-NIR spectroscopic ellipsometry (J.A. Woollam, M-2000) and was applied to the FDTD simulation file. In all FDTD simulations, we apply periodic boundary conditions along the in-plane direction. The convergence auto shut-off level was set to $10^{-5}$. The mesh size was set to 2nm for higher simulation accuracy.

*Device fabrication of Meta-PSA.* First, a layer of 130nm amorphous silicon (*a*-Si) and 60nm $SiO_x$ were deposited onto a fused silica substrate by plasma-enhanced chemical vapor deposition (PECVD) (Oxford PlasmaLab 100, 350ºC/ 15 W). Second, *a*-Si film was patterned into Si metasurface using EBL, followed by lift-off, reactive-ion etching (RIE) of SiO$_x$ mask, and ICP of *a*-Si. Third, 520nm of SiO$_x$ was sputtered onto Si metasurface to form a spacer layer. Then, a 2$^{nd}$ round of EBL alignment and patterning was performed, followed by the lift-off and RIE of the SiO$_x$ spacer layer. Next, 80nm of Aluminum was deposited onto the spacer layer using E-beam deposition, thus forming VCDGs. Finally, MPFA was diced using a wafer dicing saw, then aligned and UV-bonded onto a commercial CMOS imaging sensor (IMX477). The micro lens array and Bayer pattern layer were removed by MaxMax. corp ltd prior to the bonding process.

*Construction of Zoom lens system.* A commercial Zoom lens system for machine vision (MVL12X12Z, etc., Thorlabs Co. Ltd., USA) was applied for Zoom application. Lens attachment (MVL12X20L, Thorlabs Co. Ltd., USA) was used as objective, a ×2 extension tube (MVL20FA, Thorlabs Co. Ltd., USA), and an F-mount to C mount adapter was used to mount Meta-PSA.

*Calibration procedure for obtaining $A_{sys,T}$.* The zoom lens system was configured to ×12 times to pre-set system numerical aperture to ~0.09. We then first characterized 10 polarization states $S_{4\times10}^{in}$ generated by PSG using a Meta-PSA and register their corresponding intensity matrix $I_{6\times10}^{in}$. $A_{sys,T}$ can then be obtained by solving the equation $I_{6\times10}^{in} = A_{sys,T} \cdot S_{4\times10}^{in}$.

*Calibration procedure for obtaining $A_{sys,R}$.* The zoom lens system was configured to ×12 times to pre-set system numerical aperture to ~0.09. Next, 10 pre-known polarization states $S_{4\times10}^{in}$ reflected off a standard silver mirror (Thorlabs PF10-03-P01) were firstly generated by PSG and acquired using Meta-PSA, thus registering their corresponding intensity matrix $I_{cam,6\times10}$. $A_{sys,R}$ can then be obtained by solving the equation $I_{cam,6\times10} = A_{sys,R} \cdot M_{mirror} \cdot M_{BS,R} \cdot S_{4\times10}^{in}$. Where $M_{mirror}$ is a diagonal matrix $M_{mirror} = diag(1,1,-1,-1)$ and $M_{BS,R}$ is the combined MM of the beam splitter and objectives measured separately (see Supplementary Information section 5 for values of $M_{BS,R}$).

*Reference polarization state generation.* Light emitted from a Halogen lamp light source (Thorlabs OSL2) is first collimated using a condenser (Thorlabs ACL5040U-A) and filtered by bandpass filters with 40nm spectral bandwidth (red: FBH650-40 green: FBH500-40). Then, the collimated light is guided into either transmission or reflection light path by a flip mirror and modulated by a broadband linear polarizer (WP25M-UB by Thorlabs, Inc.) and super achromatic QWP (SAQWP05M-700 by Thorlabs, Inc.) to generate fully polarized light with arbitrary polarization states.

*Full Stokes polarization state measurement using Meta-MMM.* For transmission mode, the specimen position was left empty, i.e., $M_{s,T} = diag(1,1,1,1)$. After adding it into Eq.5, unknown input polarization states $S_{in}$ can thus be calculated by solving the equation below:

$$I_{cam} = A_{sys,T} \cdot S_{in} \qquad (12)$$

For reflection mode, a standard mirror was positioned at the specimen position, i.e., $M_{s,R} = diag(1,1,-1,-1)$. After adding it into Eq.8, unknown input polarization states $S_{in}$ can thus be calculated by solving the equation below:

$$I_{cam} = A_{sys,R} \cdot diag(1,1,-1,-1) \cdot M_{BS,R} \cdot S_{in} \qquad (13)$$

*Reference polarization state measurement with PSA using the traditional method.* Arbitrary Stokes parameters were firstly designed, then each polarization state was generated using PSG and was normally incident onto a CMOS imaging sensor with A linear analyzer (LPIREA100-C) positioned in the front. The analyzer was rotated by 2° each round, and the transmitted intensity was recorded. The recorded intensity curve was then fitted to obtain the $S_1$, $S_2$, and absolute value of $S_3$, the fitting equation is the same as Eq.S11 in Supplementary information, section 5. Then a commercial circular polarizer (Amazon Basics) was used to determine the sign of $S_3$.

*Stokes parameter measurement error analysis.* Stokes parameter measurement error is written as $\Delta S_i^j$, defined as: $\Delta S_i^j = S_i^j - S_{R_i}^j$ (i=1, 2, 3, j=1,2… 16), where $S_i^j$ is the Stokes parameter measured by our Meta-PSA and $S_{R_i}^j$ is the reference Stokes parameter. The mean absolute error (MAE) for S$_1$, S$_2$, S$_3$ is written as $\overline{\Delta_s} = \frac{\sum_{j=1,}^{16}|\Delta S_i^j|}{16}$ (i=1,2,3).

*Muller matrix of the specimen ($M_{S,T}$) calculation (transmission mode).* The solution to the Eq.8 can be written as:

$$M_{S,T} = A_{sys,T}^{-1} \cdot \begin{bmatrix} I_{P_1^1} & I_{P_1^2} & I_{P_1^3} & I_{P_1^4} & & I_{P_1^N} \\ I_{P_2^1} & I_{P_2^2} & I_{P_2^3} & I_{P_2^4} & & I_{P_2^N} \\ I_{P_3^1} & I_{P_3^2} & I_{P_3^3} & I_{P_3^4} & \cdots & I_{P_3^N} \\ I_{P_4^1} & I_{P_4^2} & I_{P_4^3} & I_{P_4^4} & & I_{P_4^N} \\ I_{P_5^1} & I_{P_5^2} & I_{P_5^3} & I_{P_5^4} & & I_{P_5^N} \\ I_{P_6^1} & I_{P_6^2} & I_{P_6^3} & I_{P_6^4} & & I_{P_6^N} \end{bmatrix} \cdot \begin{bmatrix} s_0^1 & s_0^2 & s_0^3 & s_0^4 & & s_0^N \\ s_1^1 & s_1^2 & s_1^3 & s_1^4 & & s_1^N \\ s_2^1 & s_2^2 & s_2^3 & s_2^4 & \cdots & s_2^N \\ s_3^1 & s_3^2 & s_3^3 & s_3^4 & & s_3^N \end{bmatrix}^{-1} \quad (14)$$

Notably, the requirement for a certain solution of $M_{S,T}$ is the matrix $S_{in,4\times N}$ is invertible, i.e., $rank(S_{in,4\times N}) = 4$.

*Muller matrix of specimen ($M_{S,R}$) calculation (reflection mode).* The solution to the Eq.11 can be written as:

$$M_{S,R} = A_{sys,R}^{-1} \cdot \begin{bmatrix} I_{P_1^1} & I_{P_1^2} & I_{P_1^3} & I_{P_1^4} & & I_{P_1^N} \\ I_{P_2^1} & I_{P_2^2} & I_{P_2^3} & I_{P_2^4} & & I_{P_2^N} \\ I_{P_3^1} & I_{P_3^2} & I_{P_3^3} & I_{P_3^4} & \cdots & I_{P_3^N} \\ I_{P_4^1} & I_{P_4^2} & I_{P_4^3} & I_{P_4^4} & & I_{P_4^N} \\ I_{P_5^1} & I_{P_5^2} & I_{P_5^3} & I_{P_5^4} & & I_{P_5^N} \\ I_{P_6^1} & I_{P_6^2} & I_{P_6^3} & I_{P_6^4} & & I_{P_6^N} \end{bmatrix} \cdot \begin{bmatrix} s_0^1 & s_0^2 & s_0^3 & s_0^4 & & s_0^N \\ s_1^1 & s_1^2 & s_1^3 & s_1^4 & & s_1^N \\ s_2^1 & s_2^2 & s_2^3 & s_2^4 & \cdots & s_2^N \\ s_3^1 & s_3^2 & s_3^3 & s_3^4 & & s_3^N \end{bmatrix}^{-1} \cdot M_{BS,R}^{-1} \quad (15)$$

Similarly, the requirement for a certain solution of $M_{S,R}$ is the matrix $S_{in,4\times N}$ is invertible, i.e., $rank(S_{in,4\times N}) = 4$.

*Muller matrix measurement error analysis.* The measurement error of each MM element is defined as $\Delta M_{ij}^\theta = M_{ij}^\theta - M_{R_{ij}}^\theta$ (i=0,1,2,3, j=0,1,2,3, $\theta$=60°,75°,90°,105°,120°,135°,150°), where $M_{ij}^\theta$ is the measured MM value of with the polarizer axis oriented along angle $\theta$ and $M_{R_{ij}}^\theta$ is the theoretical value of linear polarizer as a reference, written as the equation below:

$$M_R^\theta = \frac{1}{2} \times \begin{bmatrix} 1 & \cos2\theta_1 & \sin2\theta_1 & 0 \\ \cos2\theta_1 & \cos^2 2\theta_1 + \sin^2 2\theta_1 & \sin2\theta_1 \cos2\theta_1 & 0 \\ \sin2\theta_1 & \sin2\theta_1 \cos2\theta_1 & \sin^2 2\theta_1 + \cos^2 2\theta_1 & 0 \\ 0 & 0 & 0 & 0 \end{bmatrix} \tag{16}$$

The MAE of the measured MM error is defined as $\overline{\Delta_M} = \frac{\sum_{i=1,j=1,}^{4,4} |\Delta M_{ij}^\theta|}{4\times 4}$.

*Muller matrix decomposition.* Mueller matrix decomposition is based on Lu-Chipman Mueller matrix decomposition method. MM of specimen can be decomposed into a depolarizing matrix $M_\Delta$, diattenuation matrix $M_D$ and retardation matrix $M_R$:

$$M_S = M_\Delta \cdot M_R \cdot M_D \tag{17}$$

The depolarization coefficient $\Delta$ can be calculated using equation:

$$\Delta = 1 - \frac{|\text{tr}(M_\Delta) - 1|}{3} \tag{18}$$

Where tr $(M_\Delta)$ is the trace of the depolarization matrix $M_\Delta$.

The retardance $R$ can be calculated using equation below:

$$R = \text{acos}\left(\frac{\text{tr}(M_R)}{2} - 1\right) \tag{19}$$

Where tr$(M_R)$ is the trace of the retardance matrix $M_R$. Specifically, linear retardance $LR$, and circular retardance CR can then be calculated using elements of $M_R$:

$$\text{LR} = \text{acos}(sqrt\left((M_{R_{11}} + M_{R_{22}})^2 + (M_{R_{21}} + M_{R_{12}})^2\right) - 1) \tag{20}$$

$$\text{CR} = acos(\frac{M_{R_{21}} - M_{R_{12}}}{M_{R_{11}} + M_{R_{22}}}) \tag{21}$$

# Data and materials availability

All data needed to the conclusions in the paper are present in the paper and/or the Supplementary Materials.

# Reference

1. Hinrichs K, Kratz C, Furchner A. Hyperspectral mid-infrared ellipsometric measurements in the twinkling of an eye. Spectroscopy Europe. 2020.
2. Kratz C, Furchner A, Sun G, Rappich J, Hinrichs K. Sensing and structure analysis by in situ IR spectroscopy: from mL flow cells to microfluidic applications. Journal of Physics: Condensed Matter. 2020.
3. Phal Y, Yeh K, Bhargava R, editors. Polarimetric infrared spectroscopic imaging using quantum cascade lasers. Advanced Chemical Microscopy for Life Science and Translational Medicine; 2020: International Society for Optics and Photonics.
4. Stutz AJJJoAS. Polarizing microscopy identification of chemical diagenesis in archaeological cementum. 2002;29(11):1327-47.
5. Mino T, Saito Y, Yoshida H, Kawata S, Verma PJJoRS. Molecular orientation analysis of organic thin films by z‐polarization Raman microscope. 2012;43(12):2029-34.
6. He C, He H, Chang J, Chen B, Ma H, Booth MJJLS, et al. Polarisation optics for biomedical and clinical applications: a review. 2021;10(1):1-20.
7. Wang L, Zimnyakov D. Optical polarization in biomedical applications: Springer; 2006.
8. Golaraei A, Cisek R, Krouglov S, Navab R, Niu C, Sakashita S, et al. Characterization of collagen in non-small cell lung carcinoma with second harmonic polarization microscopy. 2014;5(10):3562-7.
9. Meriaudeau F, Ferraton M, Stolz C, Morel O, Bigué L, editors. Polarization imaging for industrial inspection. Image Processing: Machine Vision Applications; 2008: SPIE.
10. Sanz-Fernandez J, Saenz E, de Maagt PJIToA, Propagation. A circular polarization selective surface for space applications. 2015;63(6):2460-70.
11. Xiao H, Wang S, Xu W, Yin Y, Xu D, Zhang L, et al. The study on starch granules by using darkfield and polarized light microscopy. 2020;92:103576.
12. Lu S-Y, Chipman RAJJA. Interpretation of Mueller matrices based on polar decomposition. 1996;13(5):1106-13.
13. Jiao S, Yao G, Wang LVJAO. Depth-resolved two-dimensional Stokes vectors of backscattered light and Mueller matrices of biological tissue measured with optical coherence tomography. 2000;39(34):6318-24.
14. Ellingsen PG, Aas LMS, Hagen VS, Kumar R, Lilledahl MB, Kildemo MJJoBO. Mueller matrix three-dimensional directional imaging of collagen fibers. 2014;19(2):026002-.
15. Du E, He H, Zeng N, Sun M, Guo Y, Wu J, et al. Mueller matrix polarimetry for differentiating characteristic features of cancerous tissues. 2014;19(7):076013-.
16. Borovkova M, Trifonyuk L, Ushenko V, Dubolazov O, Vanchulyak O, Bodnar G, et al. Mueller-matrix-based polarization imaging and quantitative assessment of optically anisotropic polycrystalline networks. 2019;14(5):e0214494.

17. Wang Y, He H, Chang J, He C, Liu S, Li M, et al. Mueller matrix microscope: a quantitative tool to facilitate detections and fibrosis scorings of liver cirrhosis and cancer tissues. 2016;21(7):071112.
18. Wang Y, He H, Chang J, Zeng N, Liu S, Li M, et al. Differentiating characteristic microstructural features of cancerous tissues using Mueller matrix microscope. 2015;79:8-15.
19. Mazumder N, Qiu J, Kao F-J, Diaspro AJJoO. Mueller matrix signature in advanced fluorescence microscopy imaging. 2017;19(2):025301.
20. Ushenko YA, Sidor M, Bodnar GJQE. Mueller-matrix mapping of optically anisotropic fluorophores of biological tissues in the diagnosis of cancer. 2014;44(8):785.
21. Seitz R, Brings R, Geiger RJASS. Protein adsorption on solid–liquid interfaces monitored by laser-ellipsometry. 2005;252(1):154-7.
22. Shkirin AV, Ignatenko DN, Chirikov SN, Bunkin NF, Astashev ME, Gudkov SVJA. Analysis of Fat and Protein Content in Milk Using Laser Polarimetric Scatterometry. 2021;11(11):1028.
23. Wang W, Lim LG, Srivastava S, Bok‐Yan So J, Shabbir A, Liu QJJob. Investigation on the potential of Mueller matrix imaging for digital staining. 2016;9(4):364-75.
24. Wang C, Chen X, Chen C, Sheng S, Song L, Gu H, et al. Reconstruction of finite deep sub-wavelength nanostructures by Mueller-matrix scattered-field microscopy. 2021;29(20):32158-68.
25. Song B, Gu H, Zhu S, Jiang H, Chen X, Zhang C, et al. Broadband optical properties of graphene and HOPG investigated by spectroscopic Mueller matrix ellipsometry. 2018;439:1079-87.
26. López-Téllez J, Bruce NJAO. Mueller-matrix polarimeter using analysis of the nonlinear voltage–retardance relationship for liquid-crystal variable retarders. 2014;53(24):5359-66.
27. Freudenthal JH, Hollis E, Kahr BJCTP, Biological,, Asymmetry CCoM. Imaging chiroptical artifacts. 2009;21(1E):E20-E7.
28. Baba JS, Chung J-R, DeLaughter AH, Cameron BD, Cote GLJJobo. Development and calibration of an automated Mueller matrix polarization imaging system. 2002;7(3):341-9.
29. De Martino A, Kim Y-K, Garcia-Caurel E, Laude B, Drévillon BJOl. Optimized Mueller polarimeter with liquid crystals. 2003;28(8):616-8.
30. Laude-Boulesteix B, De Martino A, Drévillon B, Schwartz LJAo. Mueller polarimetric imaging system with liquid crystals. 2004;43(14):2824-32.
31. Arteaga O, Freudenthal J, Wang B, Kahr BJAo. Mueller matrix polarimetry with four photoelastic modulators: theory and calibration. 2012;51(28):6805-17.
32. Alali S, Vitkin IAJOE. Optimization of rapid Mueller matrix imaging of turbid media using four photoelastic modulators without mechanically moving parts. 2013;52(10):103114.
33. Huang T, Meng R, Song J, Bu T, Zhu Y, Li M, et al. Dual division of focal plane polarimeters-based collinear reflection Mueller matrix fast imaging microscope. 2022;27(8):086501.
34. Rubin NA, D'Aversa G, Chevalier P, Shi Z, Chen WT, Capasso F. Matrix Fourier optics enables a compact full-Stokes polarization camera. Science. 2019;365(6448).
35. Rubin NA, Zaidi A, Juhl M, Li RP, Mueller JB, Devlin RC, et al. Polarization state generation and measurement with a single metasurface. 2018;26(17):21455-78.
36. Tu X, McEldowney S, Zou Y, Smith M, Guido C, Brock N, et al. Division of focal plane red–green–blue full-Stokes imaging polarimeter. Applied optics. 2020;59(22):G33-G40.
37. Zuo J, Bai J, Choi S, Basiri A, Chen X, Wang C, et al. Chip-integrated metasurface full-Stokes polarimetric imaging sensor. Light: Science & Applications. 2023;12(1):218.
38. Basiri A, Chen X, Bai J, Amrollahi P, Carpenter J, Holman Z, et al. Nature-inspired chiral metasurfaces for circular polarization detection and full-Stokes polarimetric measurements. 2019;8(1):1-11.
39. Su V-C, Chu CH, Sun G, Tsai DPJOe. Advances in optical metasurfaces: fabrication and applications. 2018;26(10):13148-82.

40. Bai J, Yao YJAn. Highly efficient anisotropic chiral plasmonic metamaterials for polarization conversion and detection. 2021;15(9):14263-74.
41. Arbabi E, Kamali SM, Arbabi A, Faraon A. Full-Stokes imaging polarimetry using dielectric metasurfaces. Acs Photonics. 2018;5(8):3132-40.
42. Khorasaninejad M, Chen W, Zhu A, Oh J, Devlin R, Rousso D, et al. Multispectral chiral imaging with a metalens. 2016;16(7):4595-600.
43. Socol Y, Abramson O, Gedanken A, Meshorer Y, Berenstein L, Zaban AJL. Suspensive electrode formation in pulsed sonoelectrochemical synthesis of silver nanoparticles. 2002;18(12):4736-40.
44. Zhao Z, Chamele N, Kozicki M, Yao Y, Wang CJJoMCC. Photochemical synthesis of dendritic silver nano-particles for anti-counterfeiting. 2019;7(20):6099-104.
45. Majidi MR, Ghaderi S, Asadpour-Zeynali K, Dastangoo HJMS, C E. Synthesis of dendritic silver nanostructures supported by graphene nanosheets and its application for highly sensitive detection of diazepam. 2015;57:257-64.
46. Cheng Z-Q, Li Z-W, Xu J-H, Yao R, Li Z-L, Liang S, et al. Morphology-controlled fabrication of large-scale dendritic silver nanostructures for catalysis and SERS applications. 2019;14(1):1-7.
47. Zhou Y, Zhao G, Bian J, Tian X, Cheng X, Wang H, et al. Multiplexed SERS barcodes for anti-counterfeiting. 2020;12(25):28532-8.
48. Kozicki MNJAiPX. Information in electrodeposited dendrites. 2021;6(1):1920846.
49. Wang J, Li X, Zou Y, Sheng YJAO. Mueller matrix imaging of electrospun ultrafine fibers for morphology detection. 2019;58(13):3481-9.
50. Chen C, Ross C, Podraza N, Wronski C, Collins R, editors. Multichannel Mueller matrix analysis of the evolution of the microscopic roughness and texture during ZnO: Al chemical etching. Conference Record of the Thirty-first IEEE Photovoltaic Specialists Conference, 2005; 2005: IEEE.
51. Rezaei S, Landarani–Isfahani A, Moghadam M, Tangestaninejad S, Mirkhani V, Mohammadpoor-Baltork IJCEJ. Development of a novel bi-enzymatic silver dendritic hierarchical nanostructure cascade catalytic system for efficient conversion of starch into gluconic acid. 2019;356:423-35.
52. Bobi Olmo AR. Study of the starch grains of tubers using Mueller matrix microscopy. 2021.
53. Daly IM, How MJ, Partridge JC, Temple SE, Marshall NJ, Cronin TW, et al. Dynamic polarization vision in mantis shrimps. 2016;7(1):12140.
54. Cronin TW, Shashar N, Caldwell RL, Marshall J, Cheroske AG, Chiou T-HJI, et al. Polarization vision and its role in biological signaling. 2003;43(4):549-58.
55. Shibayev PP, Pergolizzi RGJIJB. The effect of circularly polarized light on the growth of plants. 2011;7:113-7.
56. Labhart T, Meyer EPJMr, technique. Detectors for polarized skylight in insects: a survey of ommatidial specializations in the dorsal rim area of the compound eye. 1999;47(6):368-79.
57. Edrich W, von Heiversen OJJocp. Polarized light orientation of the honey bee: the minimum visual angle. 1976;109(3):309-14.
58. Zaffar M, Pradhan AJAO. Assessment of anisotropy of collagen structures through spatial frequencies of Mueller matrix images for cervical pre-cancer detection. 2020;59(4):1237-48.

# Acknowledgments

This work was supported in part by NSF under Grant No. 2048230, 1947753, and 1809997, and DOE under Grant No. DE-EE0008999. Device fabrication and characterization in the Center for Solid State Electronics Research (CSSER) and LeRoy Eyring Center for Solid State Science (LE-CSSS) at Arizona State University was supported, in part, by NSF contract ECCS-1542160.These devices were fabricated in the Center for Solid State Electronics Research (CSSER) at Arizona State University.

# Author information

Authors and Affiliations:

**School of Electrical, Computer and Energy Engineering, Arizona State University, Tempe, AZ, USA**

Jiawei Zuo, Ashutosh Bangalore Aravinda Babu, Mo Tian, Jing Bai, Shinhyuk Choi, Hossain Mansur Resalat Faruque, Michael N. Kozicki, Chao Wang, Yu Yao

**Center for Photonic Innovation, Arizona State University, Tempe, AZ, USA**

Jiawei Zuo, Mo Tian, Jing Bai, Shinhyuk Choi , Hossain Mansur Resalat Faruque, Yu Yao

**School for Engineering of Matter, Transport and Energy, Arizona State University, Tempe, AZ, USA**

Smitha S. Swain

**Center for Molecular Design and Biomimetics at the Biodesign Institute, Arizona State University, AZ, USA**

Chao Wang

**Contributions:** Y.Y., J.Z conceived the idea, J.Z performed the theoretical analysis, J. Z. performed device simulation, J.Z., J.B. S.C., performed the device fabrication, C.W., Y.Y. supervised device fabrication, J.Z, A.B.A.B, M.T, S.B. performed device optical characterization and system calibration, A.B.A.B, H.M.R.F performed muller matrix measurement of Si wafers, J.Z, A.B.A.B, M.T performed muller matrix measurement of Si metasurface, J.Z, A.B.A.B performed muller matrix measurement of silver dendrites, J.Z. performed muller matrix measurement of honeybee wings, S.S, M.K provided the silver dendrites and SEM images, J. Z performed the data analysis. J. Z and Y.Y wrote the manuscript. All authors analyzed the results and contributed to the manuscript.

**Corresponding author:**

Correspondence to Yu Yao

# Ethics declarations

**Competing interests:** The authors declare no competing interests.